\documentstyle[preprint,revtex]{aps}
\begin{document}
\draft
\begin{title}
Central Charge of the
Parallelogram Lattice \\
Strong Coupling Schwinger Model
\end{title}
\author{Ken Yee\cite{email}}
\begin{instit}
Physics Theory Group, Brookhaven National Laboratory, Upton, NY
11973
\end{instit}
\begin{abstract}
{
We put forth a Fierzed hopping expansion
for strong coupling Wilson fermions.  As an
application, we show that the strong coupling
Schwinger model
on parallelogram lattices with nonbacktracking Wilson
fermions span, as a function of the lattice skewness
angle, the $\Delta = -1$ critical
line of $6$-vertex models.  This
Fierzed formulation also
applies to backtracking Wilson fermions, which as we describe
apparently correspond to
richer systems.  However,
we have not been able to
identify them with exactly solved models.

}
\end{abstract}
\section{Introduction and Results}
\label{intro}

	In recent years there has been remarkable progress
in the classification of two dimensional spin systems and
three dimensional topological Yang Mills theories
in relation to conformal field
theories. \cite{witten}  Yet, despite considerable activity
in lattice gauge theories with fermions \cite{wilson}
and two dimensional lattice toy models, \cite{weisz}
the Wilson lattice transcription of the Schwinger model \cite{schwinger}
remains unsolved.  The only exactly solved model with
Wilson fermions and local gauge invariance is
the square lattice
strong coupling Schwinger model, whose
partition function $Z_\Box$ at infinite hopping constant
equals $Z_{6V}[{1\over 2},{1\over 2}, 1]$, a $6$-vertex model
partition function. \cite{salmhofer}  By
known $6$-vertex model features,
this mapping reveals that $Z_\Box$
is critical and its continuum limit is
a conformal field theory with central charge $c=1$.

	Since the $6$-vertex model has a nontrivial phase
structure, it is natural to seek the $Z_\Box$
extensions which map to other regions of the
$Z_{6V}[A,B,C]$ parameter space.  To this
end, we put forth a Fierzed strong coupling hopping expansion
for actions of the form
\begin{mathletters}
\begin{equation}\label{sfdef}
S_F = \sum_{x\in\Lambda}\Bigl(
-M\overline{\psi}_x \psi_x + \sum_{\mu=0}^1
\overline{\psi}_{x+\hat{\mu}} T^{(+,\mu)} U_{x,\mu}^{\dag}
\psi_x + \overline{\psi}_{x} T^{(-,\mu)}  U_{x,\mu}
\psi_{x+\hat{\mu}}\Bigr),
\end{equation}
\begin{equation}\label{latticedef}
\Lambda\equiv\{ \sum_{\mu=0}^1 x^\mu \hat{e}_\mu
\vert x^\mu \in Z\}, ~~~
M\equiv m+ \sum_{\mu=0}^1 (T^{(+,\mu)}+T^{(-,\mu)}).
\end{equation}
\end{mathletters}
$S_F$ is the Wilson fermion action
on a square lattice action if
$\{\hat{e}_1 , \hat{e}_2\}$ are orthonormal and
$T^{(\pm,\mu )} = {1\over 2} (r\pm \gamma^\mu )$.  More generally, on a
parallelogram lattice defined by
\begin{mathletters}
\begin{equation}\label{basisdef}
\hat{e}_\mu \equiv \pmatrix{\hat{e}_\mu^{(0)}\cr
\hat{e}_\mu^{(1)}}, ~~~
\hat{e}_0 \equiv \lambda\pmatrix{ 1 \cr 0}, ~~~
\hat{e}_1 \equiv  \lambda'\pmatrix{ \cos\theta \cr
\sin\theta },
\end{equation}
\begin{equation}
g \equiv \sum_{a,b=0}^1 \hat{e}^{(a)} \hat{e}^{(b)} \delta_{ab}
=\pmatrix{\lambda^2 & \lambda\lambda'\cos\theta \cr
\lambda\lambda'\cos\theta & {\lambda}'^2},
\end{equation}
\end{mathletters}
the strong coupling partition function is
\begin{mathletters}
\begin{equation}
Z_{SC} ~\equiv ~\int_{F} ~\int
\prod_{x,\mu \in \Lambda} dU_{x,\mu} ~{\rm e\/}^{-S_F} ~,
\end{equation}
\begin{equation}
\int_{F_x}\equiv \int
d\psi_x^1 d\overline{\psi}_x^1 d\psi_x^2 d\overline{\psi}_x^2 ,
{}~~~~~~ \int_{F} \equiv \prod_{x\in\Lambda} \int_{F_x}~.
\end{equation}
\end{mathletters}

	Assuming nonbacktracking
condition
\begin{equation}\label{project1}
T^{(-,\mu)} T^{(+,\mu)} =  0 = T^{(+,\mu)} T^{(-,\mu)} ,
\end{equation}
which prevents hopping expansion
quarks from backtracking,
we will show that
\begin{equation}\label{effaction}
Z_{SC} = \int_{F}
{\rm e\/}^{-S_{SCF}} , ~~~~
S_{SCF} =  \sum_{x\in\Lambda}\Bigl(
-M\overline{\psi}_x \psi_x + \sum_{\mu=0}^1
\Theta^{(+,\mu 5)}_x \Theta_{x+\hat{\mu}}^{(-,\mu 5)} \Bigr).
\end{equation}
Commuting operators $\Theta_x^{(\epsilon,\mu 5 )}$ are
characterized by \cite{hsymmetry}
\begin{mathletters}
\begin{equation}
\int_{F_x} \bigl(\overline\psi_x\psi_x\bigr)^q
{}~\bigl(\Theta_x^{(\epsilon ,\mu 5 )}\bigr)^p =
2 \delta_{q,2} ~\delta_{p,0}
{}~~~ ~~ (\forall~q,p\in\{0,1,2,\cdots\} ),
\end{equation}
\begin{equation}\label{thetahop}
-\int_{F_x} \Theta_x^{(\epsilon' ,\nu 5)} \Theta_x^{(\epsilon,\mu 5)} =
r^{(\mu)} r^{(\nu)} \bigl(
\delta_{\mu\nu} \delta_{-\epsilon' \epsilon} +
\delta_{\mu 0}\delta_{\nu 1} S_{-\epsilon',\epsilon}^2
+ \delta_{\mu 1} \delta_{\nu 0}
S_{\epsilon, -\epsilon'}^2 \bigr),
\end{equation}
\end{mathletters}
where $r^{(\mu )}$ and $S_{\epsilon,\epsilon'}$ are functions (given
below) of $\lambda$, $\lambda'$ and $\theta$.  Since
\hbox{$\Theta_x^{(+,\mu 5)}\Theta_x^{(-,\nu 5)}$} completely
saturates $\int_{F_x}$,
$\Theta_x^{(\epsilon,\mu 5)}$ hops along selfavoiding paths.  Monomers
$(M\overline\psi_x\psi_x)^2$ fill in unhopped sites.  We call this
the Fierzed hopping expansion because the
``\hbox{$\Theta_x^{(+,\mu 5)} \Theta_{x+\hat\mu}^{(-,\mu 5)}$}''
form in (\ref{effaction})
is achieved using Fierz identities.  As described in
Section~\ref{backtracksec},
the Fierzed hopping expansion also applies to backtracking models.

	Following (\ref{effaction}),
$\Theta_x^{(\epsilon,\mu 5 )}$ hopping
amplitudes are identified in Section~\ref{vertexmod}
with Boltzmann weights of two-state vertex models.  Nonperturbative
solution of these vertex models (by Bethe ansatz or whatever) is
tantamount to resummation of the hopping expansion.  In this
way we identify
$Z_{SC}$ with $Z_{8V}$,
the $8$-vertex model partition function with Boltzmann weights
\begin{mathletters}
\begin{eqnarray}\label{startans}
\omega_1 &=&  M^2         ,~~~
\omega_2 =  0         ,~~~
\omega_3 =  (\lambda' \sin\theta)^{-2}       ,~~~
\omega_4 =  (\lambda \sin\theta)^{-2}          ,\\
\omega_5 &=& \omega_6 = {1\over 4\lambda\lambda'} \Bigl(
{1\over \cos{\theta\over 2} }\Bigr)^2 ,~~~
\omega_7 =  \omega_8 =  {1\over 4\lambda\lambda'}
\Bigl({1\over \sin{\theta\over 2} }\Bigr)^2 .
\label{endans}
\end{eqnarray}
\end{mathletters}
The $\{\omega_i\}$ are defined in, for example,
p.~$347$ of Lieb and Wu \cite{lieb}: $\omega_1$ corresponds to
the empty vertex; $\omega_3$ and $\omega_4$ to vertical
and horizontal lines; $\omega_5$ and $\omega_6$ to
lower-right and upper-left corners; and $\omega_7$ and
$\omega_8$ to lower-left and upper-right corners.  Setting
$\theta=\pi/2$ and $r^{(\mu )}=1$ recovers the
square lattice result of Ref.~\cite{salmhofer}.

	The $8$-vertex model is not
solved in general for the Boltzmann weights
given in (\ref{startans})-(\ref{endans}).  A solvable case is $M=0$ and
$\lambda ' =\lambda$.  In this subspace, define
\begin{equation}
D\equiv \omega_1=\omega_2 = 0 , ~~
C \equiv \omega_3 =\omega_4 , ~~
A\equiv \omega_5 =\omega_6 , ~~
B\equiv \omega_7 =\omega_8 .
\end{equation}
Then $8$-vertex model symmetries \cite{lieb} imply
\begin{equation}
Z_{SC}= Z_{8V} [D,C,A,B] = Z_{8V} [A,B,C,D]
= Z_{6V}[A,B,C]~.
\end{equation}
In the thermodynamic
limit, these $6$-vertex models fall on the critical line characterized
by Lieb parameter \cite{lieb}
\begin{equation}
\Delta \equiv {A^2+B^2-C^2 \over 2 A B} = -1.
\end{equation}
Varying skewness angle $\theta$ from $0$ to $\pi$ spans
the critical line between the antiferroelectric
and disorder phases of the $6$-vertex model.

	While the points on this
critical line all have central charge
$c=1$, \cite{cardy} to our knowledge it has not been demonstrated
whether they are the same $c=1$
conformal field theory or different ones.  Renormalization
group arguments suggest that the
subset of points
\hbox{$A=B={C\over 2}$}, the so-called F-model,
corresponding to
\hbox{$\theta \in \{ {\pi\over 2}, {3\pi \over 2}\}$}
can be identified with free
massless boson field theory. \cite{nienhuis}

\section{Fierzed Strong Coupling Hopping Expansion}
\label{vertexmod}

	In the finite coupling($\beta >0$) hopping
expansion \cite{wilson}
$\psi$ hops from site-to-site with weight
\hbox{$T^{(\epsilon,m )} U_{x,m}$}.  At strong coupling
such quark motions are suppressed.  Define
\begin{equation}
D_{il,jk}^{(\mu)} \equiv T^{(-,\mu)}_{ij}~ T^{(+,\mu)}_{kl} ,
{}~~~\Xi (x,y;D) \equiv
\overline{\psi}_{x}^i \psi_x^l
D_{il,jk}
\psi_{y}^j  \overline{\psi}_y^k .
\end{equation}
Since \hbox{$(\Xi)^3 =0$} in two dimensions,
integrating out $U(1)$ links $\{U_{x,\mu}\}$ using
\hbox{$ \int dU \exp(a U^{\dag} + b U ) = \sum_{k=0}^\infty
(ab)^k /(k!)^2 $} yields
\begin{mathletters}
\begin{eqnarray}\label{result}
Z_{SC} &=&\int_{F }
{\rm e\/}^{M\sum_{x} \overline{\psi}_x \psi_x }
\prod_{x,\mu }\bigl[1 + \Xi(x,x+\hat{\mu};D^{(\mu )}) +
{1\over 4}
\Xi^2(x,x+\hat{\mu};D^{(\mu )}) \bigr] \\
&=& \int_F {\rm e\/}^{\sum_x \bigl[ M\overline\psi_x \psi_x
+ \sum_\mu \bigl( \Xi(x,x+\hat{\mu};D^{(\mu )}) -{1\over 4}
\Xi^2(x,x+\hat{\mu};D^{(\mu )} ) \bigr)\bigr]}.
\label{partitionfun}
\end{eqnarray}
\end{mathletters}

	Let $n,m\in \{\pm 0,\pm 1\}$,~
\hbox{$T^{(\epsilon,-m)} \equiv T^{(-\epsilon ,m)}$},~ and
\hbox{$k_{y,-n} \equiv k_{y-\hat{n},n}$}.  Expanding
\hbox{$\exp(M\overline{\psi}_x\psi_x ) = \sum_{s_x=0}^2
(M\overline{\psi}_x \psi_x )^{s_x} / s_x !$} reveals that
$\psi\otimes\overline{\psi}$
hops with amplitude $D^{(m)}$ such that
\hbox{$s_x + \sum_{m=-1}^1 k_{x,m} = 2$}
for each \hbox{$x\in \Lambda$}.  Thus only three events are
possible at each site: (\ref{sx2})$s_x=2$,
$k_{x,m}=0$; (\ref{sk1})$s_x =1$, $k_{x,m} = 1$; or
(\ref{s0k1k1})$s_x=0$, $k_{x-\hat{n},n} = k_{x,m} = 1$.  The
associated amplitudes are
\begin{mathletters}
\begin{equation} \label{sx2}
\int_{F_x}
{1\over 2} (M \overline{\psi}_x \psi_x )^2 = M^2 ~,
\end{equation}
\begin{equation} \label{sk1}
\int_{F_x}
M \overline{\psi}_x \psi_x ~ \Xi(x,x+\hat{m} ;D^{(m )} )
= M {\rm tr\/}( T^{(+ ,m)} T^{(-,m )} \psi_{x+\hat{m}}
\overline{\psi}_{x+\hat{m}} ),
\end{equation}
\begin{equation} \label{s0k1k1}
\int_{F_x} \Xi\bigl(x-\hat{n},x;D^{(n)}\bigr) ~\Xi
\bigl(x,x+\hat{m};D^{(m)}\bigr) =
\Xi \bigl(x-\hat{n},x+\hat{m};
{\widetilde{D}}^{(n,m)}\bigr),
\end{equation}
\begin{equation}
{\widetilde{D}}^{(n,m )}_{il,jk} \equiv
(T^{(-,n )} T^{(-,m )})_{ij} (T^{(+,m )}T^{(+,n )})_{kl}-
(T^{(-,n )} T^{(+,n )})_{il} (T^{(+,m )}T^{(-,m )})_{kj}.
\end{equation}
\end{mathletters}

	A backtracking $\psi\otimes\overline\psi$ pair,
$k_{x,m}=2$ or $n=-m$ in (\ref{s0k1k1}), saturates $\int_{F}$
at the two
sites it occupies and, hence, makes a dimer.  Since
$2$-state vertex models cannot model
dimer-loop mixtures, $Z_{SC}$ is not a $2$-state vertex model
unless backtracking is forbidden.  In this
Section we adopt (\ref{project1}),
which sets $\widetilde{D}^{(n,-n)}= 0$.  Then
\hbox{$\psi\otimes\overline\psi$} worldlines
comprise a selfavoiding loop gas
with Boltzmann weight $M^2$ for unoccupied sites and weights
\hbox{$\widetilde{D}^{(n,m )} =
T^{(-,n)} T^{(-,m )}  \otimes T^{(+,m )} T^{(+,n )}$}
along loops.

	We now recast the problem so that the Boltzmann weights
are more succinctly related to hopping amplitudes.  Define
\hbox{$\gamma_5 \equiv i\gamma_0\gamma_1$},
\begin{mathletters}
\begin{equation}
\gamma_0 \equiv \pmatrix{0 & 1 \cr 1 & 0 }, ~~
\gamma_1 \equiv \pmatrix{1 & 0 \cr 0 & -1 }, ~~
\Gamma^\mu = \sum_{\nu, a}g^{\mu\nu} \hat{e}_\nu^{(a)} \gamma_a ,
\end{equation}
\begin{equation}\label{deft}
T^{(\pm ,\mu )} \equiv
{1\over 2} (r^{(\mu )} \pm \Gamma^\mu ), ~~
r^{(0 )} =  (\lambda\sin\theta )^{-1} , ~~
r^{(1 )} =  (\lambda'\sin\theta )^{-1} .
\end{equation}
Wilson regulators $\{r^{(\mu)}\}$ are chosen so that, in addition
to (\ref{project1}),
\begin{equation}\label{fierz}
T_{ij}^{(\epsilon ,\mu )} T_{kl}^{(-\epsilon ,\mu )}
= (T^{(\epsilon ,\mu )}\gamma_5)_{il}~
(T^{(-\epsilon,\mu )}\gamma_5)_{kj}~ ~~ ~(\mu~{\rm fixed\/}),
\end{equation}
\begin{equation}\label{project2}
T^{(\epsilon,\mu )} T^{(\epsilon',\mu )} = \delta^{\epsilon
\epsilon'} ~ r^{(\mu )} T^{(\epsilon,\mu )}, ~~
r^{(1)}~ T^{(\epsilon ,0)} = r^{(0 )}~
S^{\dag} ~T^{(\epsilon , 1)} ~ S ~,
\end{equation}
\begin{equation}
S=\pmatrix{\sin{\theta\over 2} & \cos{\theta\over 2}\cr
-\cos{\theta\over 2} & \sin{\theta\over 2} }\equiv
\pmatrix{S_{+,+} & S_{+,-} \cr S_{-,+} & S_{-,-}} .
\end{equation}
\end{mathletters}

	Fierz identity~(\ref{fierz}) implies
\begin{equation}\label{thetadef}
\Xi (x,x+\hat{m},D^{(m)}) = -\Theta_{x}^{(-,m 5 )}
\Theta_{x+\hat{m}}^{(+,m 5 )} , ~~~ ~
\Theta_x^{(\epsilon,\mu 5)} \equiv \overline{\psi}_x
T^{(\epsilon ,\mu )}\gamma_5 \psi_x
\end{equation}
and transforms (\ref{partitionfun}) to (\ref{effaction}).  Following
(\ref{effaction}), $\Theta_x^{(\epsilon,\mu 5)}$
hops from direction
\hbox{$n\equiv{\rm sign\/}(n) \nu$} to direction
\hbox{$m\equiv{\rm sign\/}(m) \mu$} with amplitude~(\ref{thetahop})
where \hbox{$\epsilon' ={\rm sign\/}(n)$}
and \hbox{$\epsilon = -{\rm sign\/}(m)$}.  Boltzmann
weights $\{\omega_i\}$ can be read off
from (\ref{thetahop}).  Straight $n=m$
vertical or horizontal hops, corresponding to
$\omega_3$ and $\omega_4$, have weight
$(r^{(n)})^2$.   If $x$ is approached from $n=0$ and
exited in the $m=1$ direction,
\hbox{$\omega_6 = r^{(0)} r^{(1)} S_{-,-}^2 $}.  Similarly,
$n=-1$ and $m=0$ yields
\hbox{$\omega_7 = r^{(0)}
r^{(1)} S_{+,-}^2$}.   Eqs.~(\ref{startans})-(\ref{endans}) list
the remaining vertices.

\section{Backtracking}
\label{backtracksec}

	To study backtracking effects when (\ref{project1}) is removed,
we work on a square lattice with
\hbox{$T^{(\pm, \mu )} \equiv
{1\over 2} (r\pm\gamma^\mu) $}.  Fixed-$\mu$ Fierz transformations are
\begin{equation}\label{fierzr}
T^{(\epsilon, \mu )}_{ij}
T^{(-\epsilon, \mu )}_{kl} =
(T^{(\epsilon, \mu )}\gamma_5)_{il}
(T^{(-\epsilon, \mu )}\gamma_5)_{kj}
+ \Bigl({r^2-1\over 8}\Bigr)\sum_{\sigma} (-1)^\sigma \gamma^\sigma_{il}
\gamma^\sigma_{kj},
\end{equation}
where
\hbox{$\gamma^\sigma \in \{\gamma^s\equiv 1,\gamma_5,\gamma_0,
\gamma_1\}$},
$(-1)^\sigma = -1$ for $\gamma_5$, and $(-1)^\sigma =+1$
otherwise.  Hermitian operators
\begin{equation}
\Theta^\sigma_x \equiv \cases{ \overline\psi_x \gamma^\sigma \psi_x
&if $\gamma^\sigma\in
\{\gamma^s,\gamma_5, T^{(\epsilon,\mu )}\gamma_5 \}$;\cr
i\overline\psi_x \gamma^\sigma \psi_x&if $\gamma^\sigma\in
\{\gamma_0,\gamma_1\}$;\cr}
\end{equation}
are characterized by
\begin{mathletters}
\begin{equation}
\int_{F_x} (\Theta^s_x)^2 = 2, ~~~~~
(\Theta^s_x)^3 =0,
\end{equation}
\begin{equation}
\Theta^{\sigma'}_x ~\Theta_x^\sigma = (-1)^\sigma ~
\delta_{\sigma' \sigma}~ (\Theta^s_x)^2 ~~~~ (\sigma,\sigma' \in\{s,5,0,1\}),
\end{equation}
\begin{equation}
\Theta^\sigma_x ~\Theta^{(\epsilon,\mu 5)}_x =
\Bigl[ {1\over 4} (\epsilon\epsilon'\delta_{\mu\nu} -r^2)~
\delta_{\sigma (\epsilon',\nu 5)} -{r\over 2}~
\delta_{\sigma 5} +{\epsilon \over 2} ~\epsilon^{\mu\sigma}
\Bigr] ~ (\Theta^s_x)^2 .
\end{equation}
\end{mathletters}
Following (\ref{partitionfun}) and (\ref{fierzr})
\begin{mathletters}
\begin{equation}
S_{SCF} =  S^s_{SCF} + \sum_{x,\mu
\in\Lambda} \Bigl[ \Theta_x^{(-,\mu 5)} \Theta_{x+\hat\mu}^{(+,\mu 5 )}
+\Bigl({1-r^2\over 8}\Bigr) \sum_{\sigma =5,0,1} \Theta_x^\sigma
\Theta_{x+\hat\mu}^\sigma
\Bigr],
\end{equation}
\begin{equation}
S^s_{SCF} \equiv \sum_{x\in\Lambda} \Bigl[ -M \Theta^s_x +
{1\over 32}\sum_{\mu=0}^1
\Bigl( (r^2-1) \Theta^s_x \Theta^s_{x+\hat\mu} +2\Bigr)^2 \Bigr].
\end{equation}
\end{mathletters}
Note that $S^s_{SCF}$ contains dimers.  These
systems are currently under study.  We have not
been able to identify them with solved models.

\acknowledgments
       I am indebted to Amarjit Soni and Claude Bernard for
their support, and to Yue Shen for
discussions about Ref.~\cite{weisz}.  This
manuscript has been authored
under contract number DE-AC02-76CH00016 with the
Department of Energy.

\end{document}